\documentstyle[preprint,eqsecnum,aps,epsfig]{revtex}

\begin{document}
\draft

\title{
Ultra-high energy cosmic rays: are they isotropic?
}

\author{Gustavo Medina-Tanco}

\address {
Institute of Astronomy and Geophysics,
University of Sao Paulo, Sao Paulo, Brasil \\
gustavo@iagusp.usp.br
}

\maketitle
\begin{abstract}
\widetext

From the analysis of AGASA data above $4 \times 10^{19}$ eV, we
show that the ultra-high energy cosmic rays flux is neither purely
isotropic, nor reflects the expected anisotropy from a pure source
distribution that maps large scale structure in the local universe.
The arrival distribution seems to be
the result of a mixture of fluxes (e.g., dark matter halo plus large
scale structure) or the superposition of a direct
and a diffuse radiation field components respectively. Another
viable option is an arbitrary extragalactic flux reprocessed by
a magnetized galactic wind model as recently proposed in the literature.

\end{abstract}

\vskip 1 cm

Subject headings: cosmology: dark matter --- cosmology:
large-scale structure of universe --- Galaxy: halo
--- galaxies: magnetic fields --- ISM: cosmic rays --- ISM:
magnetic fields


\narrowtext

\section{Introduction}

As ultra high energy cosmic rays (UHECR) continue to puzzle
physicists and astronomers alike, the basic question of the isotropy
of the flux beyond the unobserved GZK cut-off \cite{GZK,AG_GZK}
retains a high priority.
A recent analysis by the AGASA group ~\cite{AG_1e19} shows
a trend towards isotropy at the highest energies, although some
clusters of events (3 doublets and a triplet) are identified with
a very low chance probability. This seeming contradiction is
still more disturbing as both, bottom-up and top-down production
mechanisms generally produce anisotropy to some
degree in a natural way.
In the present letter, we address the problem of isotropy of
supra-GZK cosmic rays ($E> 4 \times 10^{19}$ eV) by using
propagation simulations and one- and two-dimensional tests
over simulated and existing world data
(mainly from AGASA). The significance of the clusters of events
detected by AGASA is also discussed in different scenarios.

\section{Numerical method and results}

Probably the conceptually simplest production models of UHECR are
the ones involving bottom-up mechanisms. All of them, require that
the sources of the particles group in more or less the same way as
luminous matter does. Furthermore, on large scales, luminous
matter trace roughly the distribution of cold dark matter (DM),
although bias factors have to be taken into account. DM is
involved in most of the top-down production mechanisms. It is
therefore important to check the expected signature from such a
source distribution (which we will call the LLMD - local luminous
matter distribution - scenario).

We use a numerical simulation approach to track UHECR propagation
through the intergalactic medium and evaluate their arrival
distribution.
The actual distribution of galaxies is used for
the UHECR sources nearer than $100$ Mpc \cite{Huchra_ZCAT}.
Additionally, the same procedure as in
\cite{GMT_Durban_IGMF97,GMT_APJL98_cluster} is used in
the description of the intergalactic magnetic field (IGMF):
a cell-like spatial structure, with cell
size given by the correlation length, $L_{c} \propto B_{IGMF}^{-2}(r)$.
The intensity of the IGMF, in turn, scales with luminous matter
density, $\rho_{gal}$ as $B_{IGMF} \propto \rho_{gal}^{0.3}(r)$
\cite{Vallee97} and the observed IGMF value at the Virgo
cluster ($\sim 10^{-7}$ G, \cite{Arp88}) is used as the normalization
condition.
Note, however, that the IGMF could be ordered and coherent on large
scales \cite{Ryu98}, in which case the propagation of UHECR
should be strongly model dependent \cite{LSS_IGMF98}.
Test particles (protons) are injected at the sources
with a spectrum $dN/dE \propto E^{-2}$ ($E < 10^{24}$ eV)
and propagated through
the intergalactic magnetic field up to the detector on Earth.
Energy losses due to redshift, pair production
and photo-pion production due to interactions with the cosmic
microwave background radiation (CMBR) are also included.

The Aitoff projection of the
resultant (2D) arrival probability density is shown in Figure 1 in
galactic coordinates.  The mask covers the plane of the galaxy,
where the actual distribution of galaxies is not well known due
to obscuration by dust. The curved, thick line is the celestial equator.
Superimposed on the figure are the available events with
$E > 4 \times 10^{19}$ eV observed by AGASA (47 events \cite{AG_1e19}),
Haverah Park (27 \cite{HP_4e19}), Yakutsk (24 \cite{YK_4e19})
and Volcano Ranch(6 \cite{VR_4e19}).
The arrival probability contours
trace quite well the local large scale structures. The supergalactic
plane (SGP), in particular, can be easily distinguished running from
North to South approximately along the $l=135$ galactic meridian.
It is apparent from the figure that, despite some conspicuous
clusters in the vicinity of the SGP, the actual observed distribution of
UHECR is much more isotropic than what one would expect if their
sources aggregate like the luminous matter. Unfortunately, given
the non-uniform exposure in declination of the various experiments
and the low number statistics involved, it is not trivial to quantify
this statement.

The view of the AGASA group \cite{AG_1e19} is that supra-GZK
events arrive isotropically at Earth. Nevertheless, to complicate
things further, three doublets and a triplet within a separation
angle of $2.5^o$ are also observed.

Independent tests are applied in order to confront AGASA data,
at the same level of statistical significance, with two opposite
yet plausible scenarios: a completely isotropic
UHECR flux and a flux originated from sources that spatially map
the large scale distribution of matter inside the GZK-sphere.



Two pools of particles, with 20 million protons each, were
constructed:
one strictly isotropic, the other obtained from the simulation
results depicted in Figure 1.
Independent samples
are extracted from these reservoirs using the response in declination
of the exposure of the AGASA experiment as a selection criteria
\cite{AG_dec_distrib}.
The size of each individual sample is equal to
the number of events (47) actually observed by AGASA above
$4 \times 10^{19}$ eV.

The most elemental analysis that can be made regarding isotropy is
one-dimensional, in right ascension (RA), where other complicating
factors like non-uniform exposure in declination and low number
statistics are more easily dealt with. Figures 2a and 2b show
different forms of visualizing the distribution of events in RA.
The shaded bands (figure 2a) in the background correspond to the
68\% and 95\% confidence levels of the expected distribution of
events in RA for a sample of size 47 protons originated in the
LLMD scenario. Despite the small size of the samples some features
are clearly seen. The largest peak is the signal from the
Virgo-Coma line of sight towards the North galactic pole. The
opposite half of the SGP (towards the second quadrant in galactic
latitude) is responsible for the smaller peak around 30$^{o}$. The
deep depressions surrounding the Virgo peak correspond to the
Orion (left) and Local (right) voids, the most prominent
structures in our immediate neighborhood, combined with the
spurious effect of obscuration of the galaxy distribution due to
the galactic plane.

The thick continuous lines in the same figure, correspond to the
68\% and 95\% confidence levels of the distribution in RA when the
incident UHECR flux is isotropic. We can see that, even with so
few events, both limits should be distinguishable.

The heavy squares represent the AGASA data (same bin size as for
the models above) and are consistent with an isotropic
distribution. No signature is seen from the Virgo peak and,
furthermore, the most populated bins fall in a region
corresponding to the Local Void.

A more quantitative treatment to characterize the anisotropy in RA
is the first harmonic analysis \cite{Linsley_1st_harmonic}. Thus,
given a data sample, the amplitude $r_{1h} = \sqrt{ a_{1h}^2  +
b_{1h}^2}$ and phase $\Psi_{1h} =\mbox{tan}^{-1} \left(
b_{1h}/a_{1h} \right)$ are calculated, where $a_{1h} = \frac{2}{N}
\sum_{i=1}^{N} \cos \alpha_{i}$, $b_{1h} = \frac{2}{N}
\sum_{i=1}^{N} \sin \alpha_{i}$ and $\alpha_{i}$ is the right
ascension of an individual event.

$r_{1h}$ and $\Psi_{1h}$ are calculated for $10^{3}$ samples drawn
form the isotropic and anisotropic (LLMD) distributions and the
results are shown in figure 2b with small dots and crosses
respectively. Both cases are very well discriminated in the
$r_{1h}$-$\Psi_{1h}$ plane. The error box for the first harmonic
of AGASA data (calculated by \cite{GMT_AAW_DM_halo}) is also
displayed (hatched region), and is completely consistent with an
isotropic UHECR flux. Moreover, the AGASA result by itself, seems
completely inconsistent with the LLMD scenario. However, when the
phase and amplitudes obtained from other major experiments are
considered (large, thick horizontal bars in Figure 2b for Haverah
Park -HP- Volcano Ranch -VR- Yakutsk -YK; see
\cite{GMT_AAW_DM_halo}) the picture looks suggestively different,
since all the phase observations are clustered inside the same
quadrant in RA, covering the right wing of the Virgo peak. That
is, despite the fact that every isolated measurement is consistent
with isotropy, the observed phases seem to show a systematic
enhancement in the direction of the interface between the SGP and
the large adjacent Local void. It must also be noted that Haverah
Park and Volcano Ranch data behave more like a transition between
the isotropic and LLMD scenarios. Three out of four first harmonic
phases (HP, YK and VR) include the North galactic pole within one
S.D. level, while the forth (AGASA) include it within two S.D..
The exclusion of the observed UHECR events inside the obscuration
band, $b < 10^{o}$, changes the phase of the AGASA result by only
$6^{o}$ (from $258^{o}$ to $252^{o}$) and, therefore, previous
conclusions are unchanged by this effect.


Clearly, a two-dimensional analysis of the data would be highly
desirably in order to answer questions as simple as whether the data
is isotropic or unimodal. One way of doing this, given the small
number of events involved and the non-uniformity of the distribution
of events in declination due to experimental limitations, is to
analyze the normalized eigenvalues
$\tau_{1}$, $\tau_{2}$ and $\tau_{3}$
of the orientation matrix $\bf{T}$ of the data. Defining
$\bf{T}_{i,j} = \Sigma_{k=1}^{N} v_{i}^{k} v_{j}^{k}$, where
$\bf{v}^{k}$ are the $N$ unit vectors representing the data over
the celestial sphere and assuming
$0 \le \tau_{1} \le \tau_{2} \le \tau_{3} \le 1$,
the shape,
$\gamma = log_{10}(\tau_{3}/\tau_{2})/log_{10}(\tau_{2}/\tau_{1})$,
and the strength parameter,
$\zeta = log_{10}(\tau_{3}/\tau_{1})$,
can be built \cite{Stats_sphere}.
The shape criterion $\gamma$ is useful in discriminating
girdle-type distributions from clustered distributions. The larger
the value of $\gamma$ more clustered is the distribution. Uniform,
nearly isotropic, distributions have $\zeta \sim 0$. Because of the
nature of the experimental setup, the observed distribution of
UHECR is girdle in nature,
regardless of the isotropicity of the UHECR flux.
Therefore, in figure 3 we compare the results for $10^{3}$
isotropic (rhombes) and LLMD (circles) samples respectively
with the AGASA sample
in the $\gamma$-$\zeta$ plane. It can be seen that the
isotropic and LLMD scenarios should be very well separated
with the available data, albeit its smallness.
AGASA data (thick cross), on the other hand, does not fit
either of these scenarios, being an intermediate case.

Figure 4 shows the number of doublets with separation smaller than
2.5$^{o}$ obtained from $10^{4}$ samples of $47$ events each,
drawn from isotropic and LLMD populations. AGASA observed 3 pairs,
which is a large number (~8\% chance probability) for an isotropic
UHECR flux, and a rather small number (but still inside the 68\%
C.L. at ~13\% probability) if compared with the average value of
~5.5 pairs obtained for the anisotropic flux. The situation is
analogous for triplets. One triplet was actually observed by
AGASA, while $1.3 \pm 3.1$ is expected for the LLMD model and
$0.02 \pm 0.2$ for the isotropic model. It should also be noted
that the AGASA triplet C2 and pair C1 (actually a triplet if
Haverah Park data is included), as well as the lower energy
cluster BC2 fall on the SGP, on top of a maximum of the arrival
probability \cite{GMT_APJL98_cluster}, strengthening the case for
an extragalactic origin inside this structure.


\section{Discussion and implications}

Different tests have been applied to the analysis of the
isotropy of UHECR with $E>4 \times 10^{19}$ eV. Test samples
are drawn from both, an isotropic flux of particles and
an anisotropic flux originated in sources with the local
luminous matter distribution. Samples used to compare with
AGASA data have a size of 47 events and are selected according
to the same declination sensitivity as AGASA's.
Our results can be summarized as follows:

1) The comparison between arrival directions of the UHECR
above $4 \times 10^{19}$ eV and the expected arrival
probability density, calculated under the assumption of UHECR
sources that cluster as the luminous matter in the nearby
universe, shows a remarkable degree of isotropy, despite a
notorious tendency for clusters to appear on top of large
scale structure signatures (Fig.1).

2) AGASA's RA distribution is consistent with an isotropic
distribution (Fig.2).

3) $10^{4}$ simulated experiments equivalent to AGASA show that a set as
small as 47 UHECR is enough to separate extremely well isotropic
and LLMD scenarios in the amplitude-phase plane; AGASA's error
box is completely consistent with isotropy and inconsistent
with LLMD (Fig.2)

4) Nevertheless, the phases of HP, VR, YK and AGASA fall in the
same quadrant in phase (Fig.2), which covers the interface
between the SGP  in the general Virgo direction
and the adjacent Local void. The first three experiments include
the North galactic pole inside 1 SD, and AGASA at 2 SD.

5) The comparison of AGASA data with $10^{4}$ simulated data
sets from isotropic and LLMD fluxes on the $\gamma$-$\zeta$
(shape-strength) plane show that the former is an intermediate
case, more clustered than isotropic samples but less than LLMD
(Fig.3).

6) The number of pairs observed by AGASA is too large for an
isotropic flux, but it is within the 68\% CL for LLMD
flux (Fig.4).

While a first order interpretation of the AGASA data certainly
points to an isotropic flux of UHECR, consideration of the first
harmonic analysis of other data sets and of two-dimensional
tests over the AGASA data itself, as well as expected numbers
of doublets for isotropic and anisotropic samples, point to a
more complicated, intermediate picture with a certain degree
of mixture of both limiting cases.

We can envisage at least three scenarios in which such a
result could be obtained:

1) The sources involve bottom-up mechanisms associated with
luminous matter but some of the events are scattered in the
intergalactic medium such that we observe the composition
of a diffuse and a direct component \cite{ghosts}.

2) The sources involve bottom-up mechanisms associated with
luminous matter but there is a large local magnetic structure,
like a magnetized galactic wind, which isotropize the UHECR flux
upon traversing the galactic halo \cite{gal_wind}. As the
energy of the particles increases, and as long as they all have
the same mass, the degree of isotropization should decrease
making the galactic pole visible.

3) The sources involve top-down mechanisms associated with
dark matter whose distribution roughly associates with the LLMD.
In this case, the observed flux is the composition of an
extragalactic component, whose signature is not very different from
that of the LLMD, and a component originated in the halo of our
own galaxy. \cite{Dubovsky_DM} showed that, under
general conditions, the halo component would dominate the
extragalactic flux by at least two orders of magnitude. This is
only true, however, in the unrealistic case of dark matter
uniformly distributed in intergalactic space. Nevertheless,
dark matter aggregates strongly and tends to be overabundant,
by factors of $\sim 10^{2}$, in the center of galaxy clusters when
compared to its abundance in the halos of isolated galaxies.
It can therefore be shown that, in a sample of 47 events,
and assuming Virgo as the only source of extragalactic events,
3-7 events should originate in Virgo and arrive inside a solid angle
of approximately the size of the cluster. This could give rise
to a slight anisotropy that correlates with the SGP
when combined with the almost isotropic flux originated
in a large galactic halo. Note, however, that the solid angle
does not need to point exactly in the direction of Virgo,
depending on the large scale structure of the intervening
magnetic field.

Obviously, more high quality, high energy data from HiRes
and Auger are badly needed.

{\bf Acknowledgments.}
The author benefited from discussions with Alan. A. Watson,
Peter Biermann, Todor Stanev, Eun-Joo Ahn and Torsten Ensslin.
This work is partially supported by the Brazilian agencies
FAPESP and CNPq.


\begin{figure}[!hbt]
\caption{ Arrival probability distribution for the LLMD model
(contour lines in the background) compared with the actual
published data above $4 \times 10^{19}$ eV from AGASA, Haverah
Park, Yakutsk and Volcano Ranch. \label{lbl_fig1}}
\end{figure}

\begin{figure}[!hbt]
\caption{ (a) Arrival directions (right ascension) of UHECR above
$4 \times 10^{19}$ eV for the LLMD (hatched regions) and isotropic
(lines) models, as well as the observed AGASA data (squares). 68\%
and 95\% confidence levels are shown for both models. Confidence
levels are calculated for each individual bin after 1000
independent experiments (with 47 events each - same number as the
AGASA sample) were performed. (b) Amplitude and phase of the first
harmonic calculated for $10^{3}$ samples drawn form the isotropic
(circles) and anisotropic (LLMD - crosses) distributions. The size
of individual samples is 47 protons, as in AGASA. The hatched
region is the ($1 \sigma$) error box calculated from AGASA
observations, while the thick horizontal bars are the $1 \sigma$
error bars for the phases of Volcano ranch, Haverah Park and
Yakutsk experiments. In both cases, (a) and (b), samples are
selected with the same declination distribution expected for the
AGASA experiment. \label{lbl_fig2}}
\end{figure}

\begin{figure}[!hbt]
\caption{Two-dimensional eigenvector analysis (see the text for
the definition of the shape and strength parameters). The heavy
cross is the AGASA observation. Rhombuses and circles correspond
to isotropic and LLMD simulations respectively. \label{lbl_fig3}}
\end{figure}

\begin{figure}[!hbt]
\caption{Expected frequency of doublets for the isotropic
(crosses) and LLMD (rhombuses) models respectively. The horizontal
bars show the expected mean and one standard deviation intervals
for each model. \label{lbl_fig4}}
\end{figure}


\newpage

\begin{figure}[!hbt]
\centerline{\psfig{figure=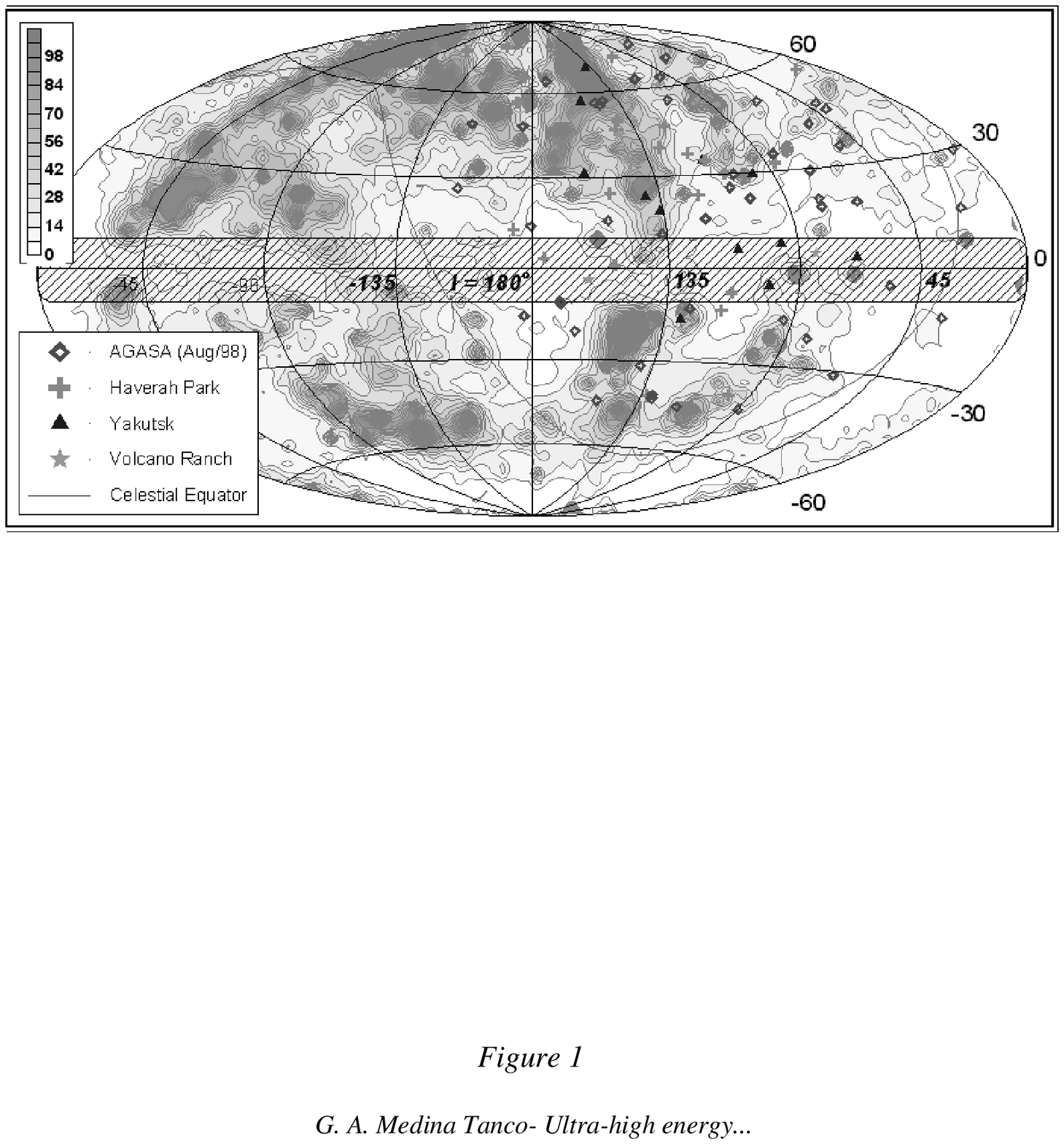}}
\end{figure}

\begin{figure}[!hbt]
\centerline{\psfig{figure=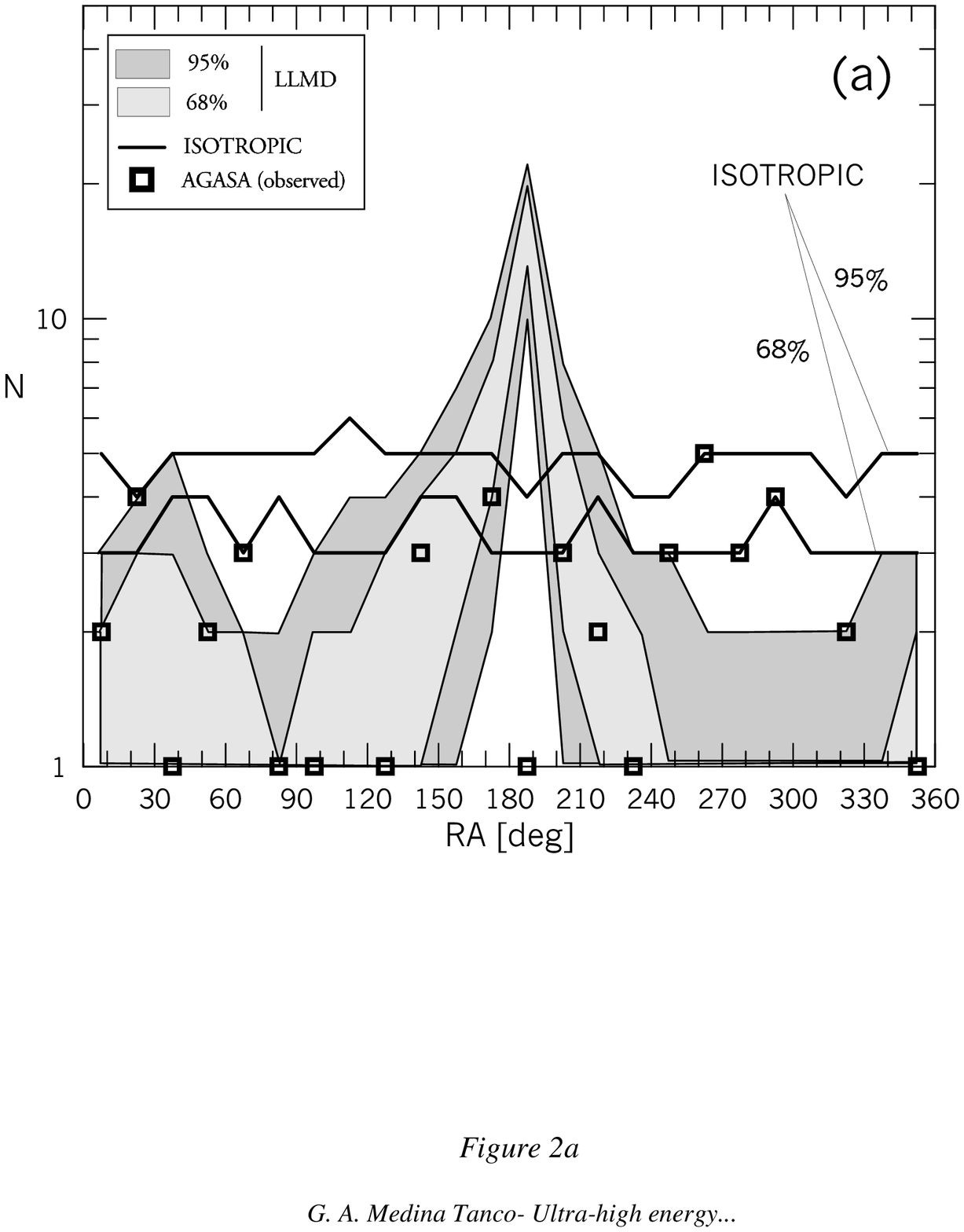,width=16cm}}
\end{figure}

\begin{figure}[!hbt]
\centerline{\psfig{figure=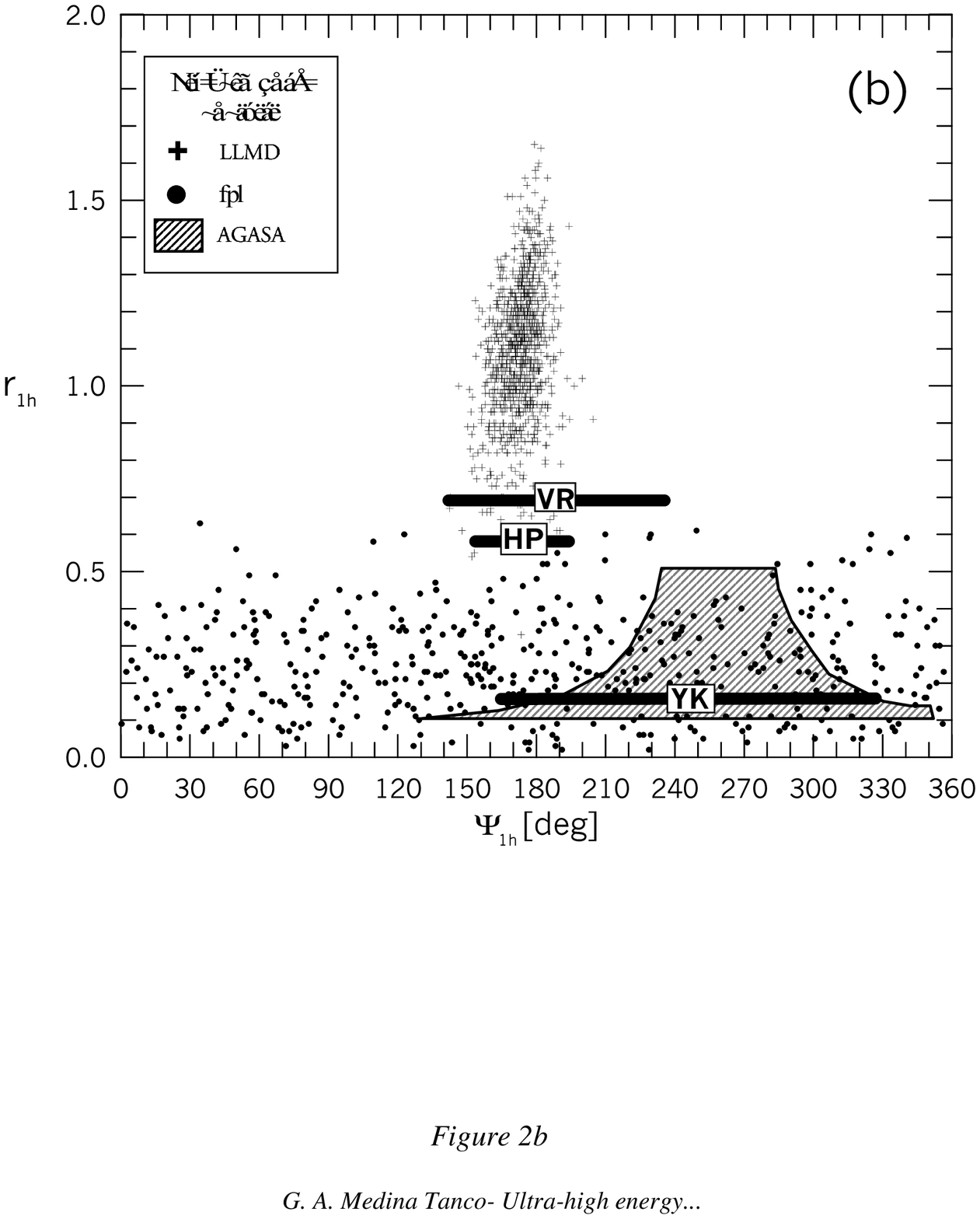,width=16cm}}
\end{figure}

\begin{figure}[!hbt]
\centerline{\psfig{figure=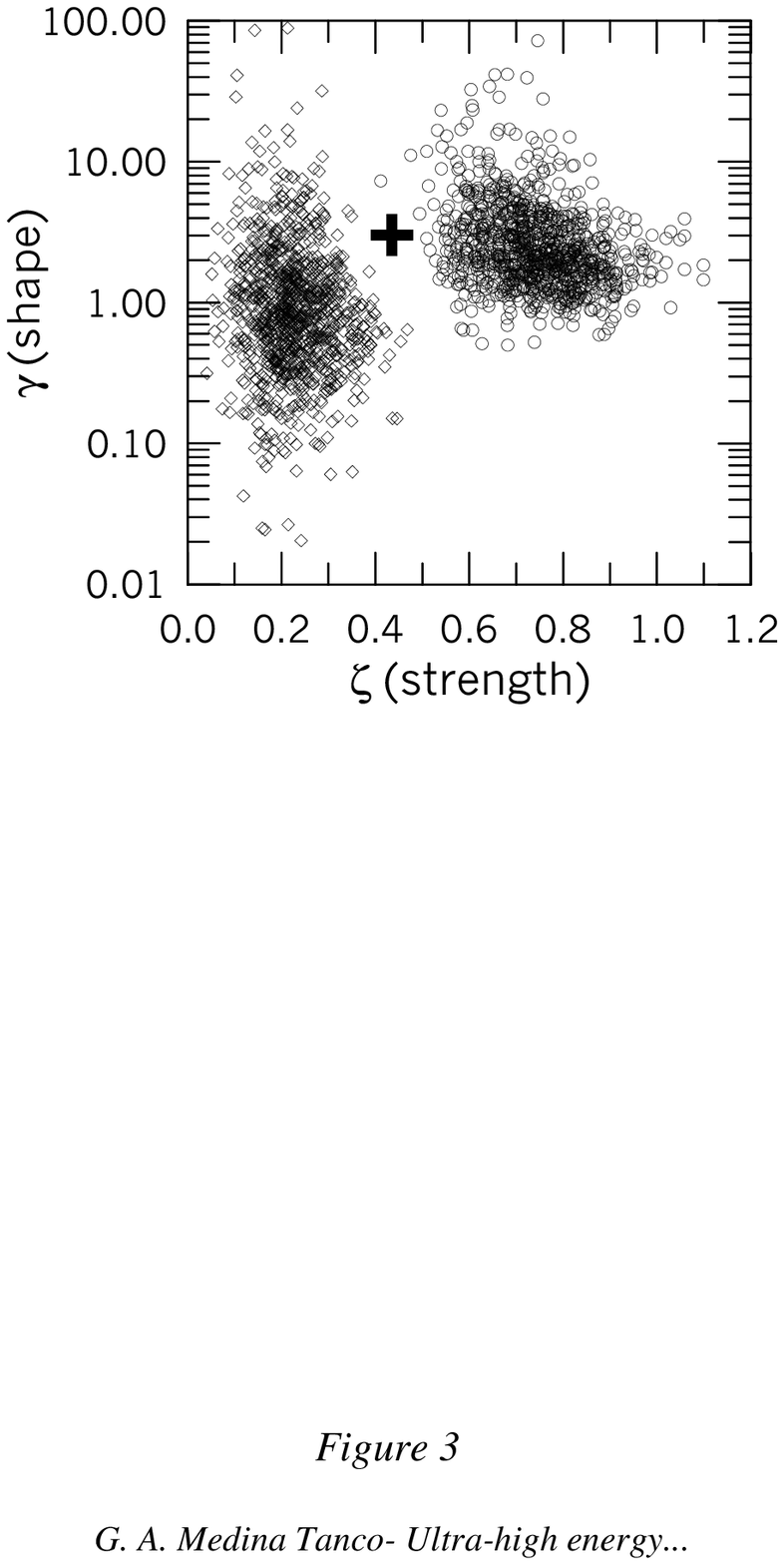,width=16cm}}
\end{figure}

\begin{figure}[!hbt]
\centerline{\psfig{figure=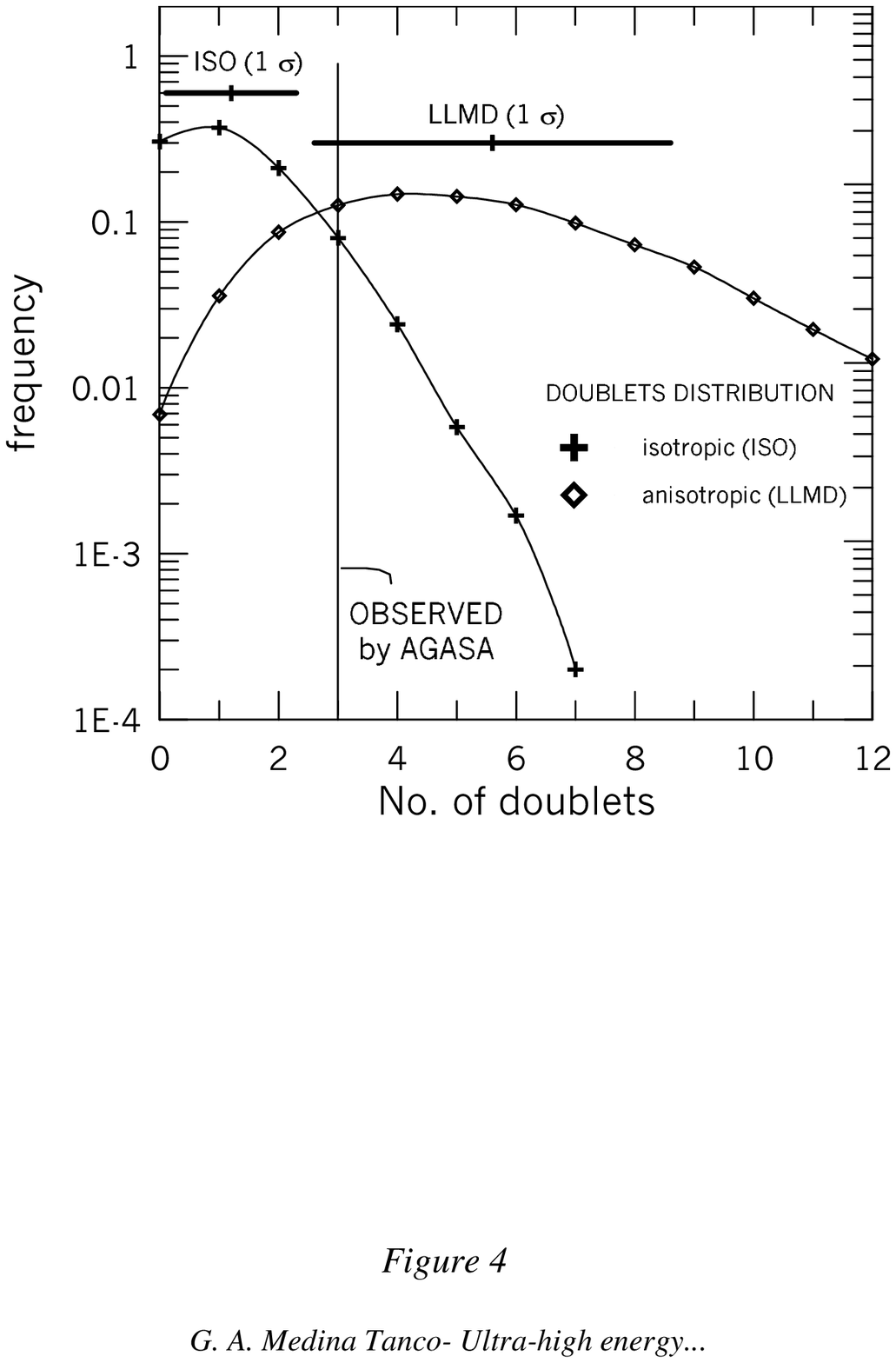,width=16cm}}
\end{figure}

\end{document}